\documentclass[pra,floatfix,twocolumn]{revtex4}

\usepackage{amssymb,graphics,graphicx,times,bm,multirow}

\usepackage{times,amsmath,amsfonts,amssymb,latexsym}
\usepackage{graphicx,epsf,epsfig,verbatim}

\usepackage{epstopdf}
\usepackage{setspace}
\usepackage{amsmath}
\usepackage{float}
\usepackage{color}
 \usepackage{bbold} 

\newtheorem{theorem}{Theorem}

\begin{document}

\title{Artificial quantum thermal bath:\\ Engineering temperature for a many-body quantum system}

\author{Alireza Shabani and Hartmut Neven}

\affiliation{Google Inc., Venice, CA, 90291}

\date{\today}

\begin{abstract}
Temperature determines the relative probability of observing a physical system in an energy state when that system is energetically in equilibrium with its environment. In this paper, we present a theory for engineering the temperature of a quantum system different from its ambient temperature. We define criteria for an engineered quantum bath that, when coupled to a quantum system with Hamiltonian $H$, drives the system to the equilibrium state $\frac{e^{-H/T}}{{{\rm{Tr}}}(e^{-H/T})}$ with a tunable parameter $T$.  This is basically an analog counterpart of the digital quantum metropolis algorithm. For a system of superconducting qubits, we propose a circuit-QED approximate realization of such an engineered thermal bath consisting of driven lossy resonators. Our proposal opens the path to simulate thermodynamical properties of many-body quantum systems of size not accessible to classical simulations. Also we discuss how an artificial thermal bath can serve as a temperature knob for a hybrid quantum-thermal annealer.
\end{abstract}
\maketitle

\section{Introduction}

How can we change the temperature of a physical system? A quick answer is by changing the temperature of its surrounding environment. For many applications this is an impractical option. For instance, a superconducting quantum processor is kept in a fridge at a few millikelvin. Raising the temperature of the fridge would excite various sources of noise annihilating the quantumness of the system. This article proposes an alternative solution: {\it{an artificial temperature}}. A tunable temperature allows physical simulation of thermodynamics of many-body quantum systems \cite{Georgescu:14}, and it also serves as a computational knob for a processor performing the annealing method of optimization \cite{Annealing:Book,Nishimori:Book}.
  
Calculating the thermal state or partition function of a many-body physical system is a fundamental computational problems in and a bridge across statistical physics \cite{Pathria:Book}, quantum computation \cite{Annealing:Book,Nishimori:Book}, and machine learning \cite{Bishop:Book}. In statistical physics, estimating the partition function of a many-body quantum system is known to be computationally hard \cite{Newman, Troyer:2005}. Quantum metropolis algorithms have been proposed as a solution, although they demand a fault-tolerant universal digital quantum computer \cite{Terhal:02,Poulin:09,Temme:11,Yung:12,Riera:12}. Therefore, it is of paramount importance to realize quantum simulators to generate the thermal equilibrium state of physical systems, $\frac{e^{-H_S/T}}{{{\rm{Tr}}}(e^{-H_S/T})}$. Such a simulator should have two features: a programmable Hamiltonian $H_S$ and a tunable temperature $T$.

Another paradigm that requires a tunable temperature is an analog annealing processor. Annealing is a general purpose method of optimization where the solution of a computational problem is encoded in the ground state of a system Hamiltonian $H_S$. System starting in an excited state is gradually guided to lower energy states and eventually to the ground state which carries the optimal solution. Classical annealing uses thermal fluctuations to explore the optimization landscape and the energy is reduced by lowering the temperature \cite{Nishimori:Book}. Therefore, a Gibbs state simulator can in principle be used as a classical annealer when the temperature $T(t)$ is time-dependent. A zero-temperature paradigm of annealing, known as quantum annealing, employs quantum fluctuation to drive the system toward the ground state. However, there is theoretical evidence that annealing at non-zero temperature or in the presence of noise can be computationally more powerful than a zero temperature quantum annealer \cite{Brooke:99,Amin:08,Tanaka:11,Smelyanskiy:15,Kechedzhi:16,Nishimura:16}. These findings suggests that future quantum annealing processors should be equipped with a temperature knob, basically making them a hybrid quantum-classical annealer.

Here we present a realization of an artificial temperature by quantum bath engineering. We reference the method of cavity cooling, which addresses a similar question for a single mode harmonic oscillator \cite{Clerk:10,Teufel:11}, a system of non-interacting qubits \cite{Kapit:2015}, and a Bose-Hubbard chain with known energy structure \cite{Hacohen:2015}. Here we avoid any hypothetical assumption such as bath eigenstate thermalization in Ref. \cite{Fialko:2015}. Quantum bath engineering, as a powerful resource for quantum engineering, has been recently applied to other applications such as quantum state preparation \cite{Kraus:08, Murch:12, Shankar:13} and realizing a chemical potential for photons \cite{Hafezi:14}. In the following, we first define the notion of a universal thermal bath and derive the microscopic conditions for engineered thermalization. Then we give a detailed proposal for realization of such a bath for superconducting qubits.

\section{Universal Quantum Thermal Bath}

 We begin with an {\it{operational}} definition of temperature since we consider temperature as a knob for an analog computation. Given a quantum system $S$ with Hamiltonian $H_S$, we say the system is at temperature $T$ if its state is represented by the density matrix $\frac{e^{-H_S/T}}{{{\rm{Tr}}}(e^{-H_S/T})}$. Correspondingly, we define a {\it{universal}} artificial quantum thermal bath as a quantum system $B$ that, when coupled to the system $S$, sets its temperature with the following criteria\\

{\it{(i) The bath Hamiltonian and system-bath interaction are independent of the system Hamiltonian.}}\\

{\it{(ii) The bath internal dynamics is tunable.}}\\

{\it{(iii) The strength of system-bath interaction is tunable.}}\\

Condition (i) carries the notion of universality by ensuring that as the system Hamiltonian changes (e.g. programming problem Hamiltonian of an annealer) the bath thermalization functionality is intact. Conditions (ii) and (iii), as described below, correspond to tunability of the temperature and rate of thermalization.

\subsection{Microscopic Theory}

The theory of open quantum systems offers a detailed description of how quantum systems thermalize when coupled to a {\it{natural}} thermal bath \cite{Breuer:Book}. Here the word natural refers to the surrounding environment and distinguishes it from an engineered bath. In this case, the bath itself is assumed to be at temperature $T_{env}$ in state $\frac{e^{-H_B/T_{env}}}{{{\rm{Tr}}}(e^{-H_B/T_{env}})}$. If the bath is weakly and linearly coupled to the system and its fluctuations satisfy the Kubo-Martin-Schwinger (KMS) condition , the equilibrium state of the system is guaranteed to be $\frac{e^{-H_S/T_{env}}}{{{\rm{Tr}}}(e^{-H_S/T_{env}})}$ \cite{Breuer:Book}, obeying thermodynamic laws. A natural thermal bath is not universal as there is no switch to turn it off. 

 We develop a theory for an analog quantum thermal bath based on engineering the bath's dynamical fluctuations and KMS conditions. We begin with the derivation of a quantum master equation for a bath out of thermal equilibrium with time-dependent system-bath interaction. We consider system $S$ as weakly coupled to an auxiliary quantum system (bath) $B$. In superconducting systems, the bath can consist of resonators \cite{Blais:04, Erickson:14, Grezes:14, Sundaresan:15}, qubits, or meta-materials \cite{Egger:13,Haeberlein:15,Plourde:15}. For a total system-bath Hamiltonian $H_{SB}=H_S+H_B+H_I(t)$, dynamics are described by 
\begin{equation}
\frac{d\tilde{\rho}_{SB}}{dt}=-i[\tilde{H}_I(t), \tilde{\rho}_{SB}(t)]
\end{equation}
where $\sim$ refers to the interaction picture.

For weak system-bath interaction, the second-order perturbative solution of the dynamics yields
\begin{eqnarray}
\frac{d\tilde{\rho}_S}{dt}&=&-i\mbox{tr}_B [\tilde{H}_I(t), \rho_{SB}(0)]\notag\\
&-&\int_0^t ds \mbox{tr}_B[\tilde{H}_I(t),[\tilde{H}_I(s),\tilde{\rho}_{SB}(s)]]
\end{eqnarray}
Here we neglect the first term on the RHS, assuming $tr_B [\tilde{H}_I(t), \rho_{SB}(0)]=0$. We will discuss this assumption in section II.C.

The next assumption we use is the Markovian approximation. For this we need the auxiliary system $B$ to be strongly attracted to an equilibrium state $\rho_B^{SS}(t)$ such that, after any perturbative kick by the system, it quickly relaxes back to its equilibrium state. In general, the bath steady-state can be time dependent. An example of such an auxiliary system that we propose below is a strongly lossy resonator in a coherent equilibrium state. As a result of the Born-Markov approximation we have $\tilde{\rho}_{SB}(t)\approx \tilde{\rho}_{S}(t)\otimes \tilde{\rho}_B^{SS}(t)$ and arrive at

\begin{equation}
\frac{d\tilde{\rho}_S}{dt}=-\int_0^\infty ds \mbox{tr}_B[\tilde{H}_I(t),[\tilde{H}_I(t-s),\tilde{\rho}_{S}(t)\otimes \tilde{\rho}_B^{SS}(t)]]\label{redfield}
\end{equation}

Next we consider system-Bath interaction terms of the form $H_I=\sum_\alpha g_\alpha S_\alpha\otimes B_\alpha(t)$ (with Hermitian operators $S_\alpha$ and $B_\alpha$). Given projective decomposition $H_S=\sum_\epsilon \epsilon \Pi(\epsilon)$, we define the system operators as
\begin{equation}
S_\alpha(\omega)=\sum_{\epsilon'-\epsilon=\omega}\Pi(\epsilon)S_\alpha\Pi(\epsilon')
\end{equation}
We apply the rotating-wave approximation by constraining the coupling strengths $g_\alpha$ such that the resulting thermalization rate is weaker than the minimum system frequency $\Delta_S=\min (\epsilon'-\epsilon)$.  Applying this approximation to (\ref{redfield}), we obtain the Lindblad equation
\begin{eqnarray}
&&\frac{d\tilde{\rho}_S}{dt}=\mathcal{D}(\tilde{\rho}_S)= \label{Lindblad}\\
&&\sum_{\omega,\alpha,\alpha'}\gamma_{\alpha\alpha'}(\omega,t)\Big(S_{\alpha'}(\omega)\tilde{\rho}_S S^\dagger_\alpha(\omega)-\frac{1}{2}\{S^\dagger_\alpha(\omega)S_{\alpha'}(\omega),\tilde{\rho}_S\}\Big)\notag
\end{eqnarray}
with fluctuation coefficients
\begin{eqnarray}
\gamma_{\alpha\alpha'}(\omega,t)=g_\alpha g_{\alpha'}\int_{-\infty}^\infty ds e^{i\omega s} {\mbox{tr}}(B_\alpha(t)B_{\alpha'}(t-s)\rho_B^{SS}(t))
\end{eqnarray}

The Born approximation translates to $|\gamma_{\alpha\alpha'}|\ll \Delta_S, \gamma_B$ where $\Delta_S$ and $\gamma_B$ are the system's minimum energy gap and bath relaxation rate, respectively. In the case of annealing dynamics such that the system Hamiltonian $H_S(t)$ varies over time, we need the further assumption of slow annealing to apply the master equation in terms of the instantaneous eigen-energies, $H_S(t)=\sum_\epsilon \epsilon_t \Pi(\epsilon,t)$ \cite{Albash:12}. Next we discuss how to engineer the bath fluctuations to realize a universal thermal bath.

\subsection{Thermalization Condition}

The following relation is a {\it{sufficient}} condition on the dynamics (\ref{Lindblad}) having the Gibbs state $\rho_{th}=\frac{e^{-H_S/T}}{{\rm{Tr}}(e^{-H_S/T})}$ as the steady-state,

\begin{equation}
\gamma_{\alpha\alpha'}(-\omega,t)=e^{-\omega/T}\gamma_{\alpha'\alpha}(\omega,t) \label{thermal_rondition}
\end{equation}
For time-homogenous fluctuations where $\gamma_{\alpha\alpha'}$ is time-independent, the uniqueness of the steady-state of (\ref{Lindblad}) guaranteed if any two energy levels $\{|\epsilon\rangle,|\epsilon'\rangle\}$ are coupled either directly through an operator $S_\alpha$ ($\langle\epsilon|S_\alpha|\epsilon'\rangle\neq 0$) or indirectly via intermediate levels $\{|\epsilon_j\rangle\}$ ($\langle\epsilon|S_j|\epsilon_j\rangle ... \langle\epsilon_{j'}|S_{j'}|\epsilon'\rangle\neq 0$). See appendix A for details of the thermalization condition. Condition (\ref{thermal_rondition}) is a generalization of the KMS condition, which is automatically satisfied if the bath steady-state is thermal, $\rho_B^{SS}=\frac{e^{- H_B/T}}{{\rm{Tr}}(e^{- H_B/T})}$ \cite{Breuer:Book}. The Lindblad generator (\ref{Lindblad}) that satisfies the thermalization condition is known as the Davies map \cite{Davies:79,Temme:13}. We next discuss how to engineer the KMS condition (\ref{thermal_rondition}) with a non-thermal bath.

\subsubsection{Approximate Thermalization Condition}
\begin{figure}[h]
\includegraphics[width=6cm,height=4.5cm]{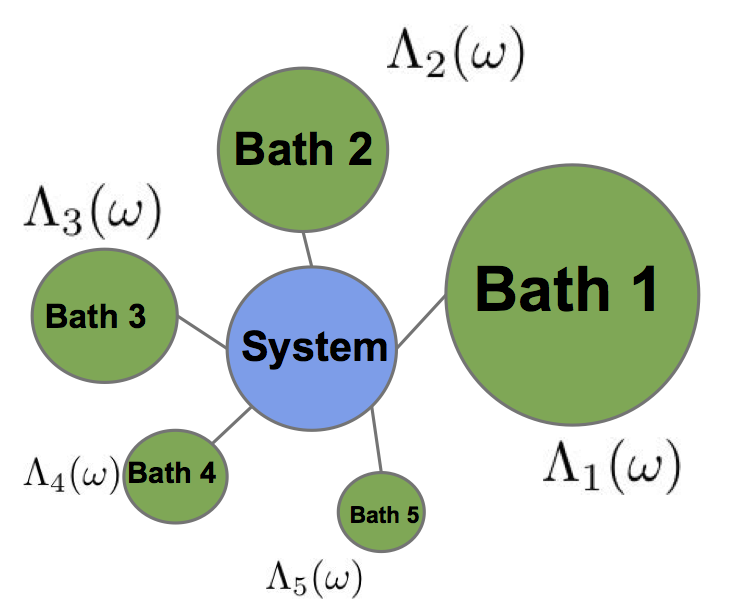}
\caption{A cartoon of a thermal bath, a discrete set of baths that collectively satisfy the KMS condition (\ref{thermal_rondition}).}\label{resonator_bath}
\end{figure}

Here we propose one approach to engineering the KMS condition by using a collection of engineerable baths (fig. \ref{resonator_bath}). Consider bath modes $B_\nu$ coupled to the system via the interaction Hamiltonian $H_I=\sum_{\alpha} S_\alpha\otimes \sum_{\nu} g_{\alpha,\nu}B_{\nu}$ with time-homogenous bath fluctuations 
\begin{eqnarray}
\Lambda_{\nu\nu'}(\omega)= \int_{-\infty}^\infty ds e^{i\omega s} {\rm{Tr}}(\tilde{B}_{\nu}(s)\tilde{B}_{\nu'}(0)\rho_{\nu}^{SS})
\end{eqnarray}
We engineer baths such that collectively satisfy condition (\ref{thermal_rondition}) to a certain precision:
\begin{eqnarray}
&\gamma_{\alpha\alpha'}(\omega)=\sum_{\nu,\nu'}g_{\alpha,\nu}g_{\alpha',\nu'}\Lambda_{\nu\nu'}(\omega) \notag\\
 &{\mbox{ such that }}   \gamma_{\alpha\alpha'}(-\omega)\approx e^{-\omega/T}\gamma_{\alpha'\alpha}(\omega) \label{KMS_approx}
\end{eqnarray}
with $\rho_{\nu}^{SS}$ is the $\nu$'th bath steady-state. This condition should hold in a relevant range of energies $[\omega_{min},\omega_{max}]$. 

Next we analyze the required precision of this engineering approach as any engineered KMS condition would inevitably be an approximation to condition (\ref{thermal_rondition}). 


\subsubsection{Precision Analysis}
An approximate condition (\ref{KMS_approx}) on bath fluctuations guarantees an approximate preparation of a thermal state. We make this statement explicit by proving the following bound on the distance between a target thermal state $\rho_{th}$ and the actual engineered equilibrium state $\rho_{eq}$. Consider a system with Hilbert space dimension $d$ coupled to a bath with engineered fluctuations $\gamma_{\alpha\alpha'}(\omega)$, then we have

\begin{eqnarray}
||\rho_{eq}-\rho_{th}||_1&\leq 6\frac{\log(d)+1}{\lambda} G^2(d)\times\notag\\
\max_{\alpha,\alpha',\omega\in [-\omega_{max},-\omega_{min}]}&\Big| e^{\omega/T}\gamma_{\alpha'\alpha}(-\omega)-\gamma_{\alpha\alpha'}(\omega)\Big|\label{precision}
\end{eqnarray}
where for an arbitrary system Hamiltonian $G(d)=\frac{1}{2}d(d+1)$, however for an Ising Hamiltonian $G(d)=\frac{1}{2}(\log^2(d)+\log(d))$ that means quadratic with the number of qubits. The factor $\lambda$ is the gap of Davies map that we introduce in appendix B to derive inequality (\ref{precision}). The Davies map corresponding to the Lindblad superoperator (\ref{Lindblad}) is defined by fluctuation functions $\gamma^*_{\alpha\alpha'}(\omega)=\gamma_{\alpha\alpha'}(\omega)$ for $\omega\geq 0$, and $\gamma^*_{\alpha'\alpha}(\omega)=\gamma_{\alpha'\alpha}(-\omega)e^{\omega/T}$ for $\omega<0$. The parameter $\lambda$ is the first non-zero eigenvalue which is of course negative and determines the rate of thermalization. The distance is measured by trace norm $||.||_1$. See appendix B for the derivation of the bound (\ref{precision}).

The inequality (\ref{precision}) is a powerful tool that relates the accuracy of preparing a target Gibbs state to the precision in the bath engineering. Basically if a design requires accuracy $\epsilon$, i.e. $||\rho_{eq}-\rho_{th}||_1\leq \epsilon$ that translates into engineering bath fluctuations with precision
\begin{eqnarray}
\Big| e^{\omega/T}\gamma_{\alpha'\alpha}(-\omega)-\gamma_{\alpha\alpha'}(\omega)\Big|\leq \frac{\epsilon\lambda}{6(\log(d)+1)G^2(d)}
\end{eqnarray}
for $-\omega_{max}\leq\omega\leq-\omega_{min}$. This bound can become explicit further by using the results of Ref. \cite{Temme:13} on lower bounds for the gap $\lambda$ based on the elements of the superoperator (\ref{Lindblad}).

\subsection{Initial Conditions for Simulation and Annealing Modes} In the early steps of deriving master equation (\ref{redfield}), we assumed the following initial state condition
\begin{equation}
tr_B [\tilde{H}_I(t), \rho_{SB}(0)]=tr(\tilde{B}(t) \rho_{B}(0)) [\tilde{S}(t), \rho_{S}(0)]=0, \label{initial_state}
\end{equation}
which can be guaranteed in the following two ways:\\

I) Quantum simulation mode: Here the goal is to generate a Gibbs state for a time-independent Hamiltonian $H_S$. In this case the condition (\ref{initial_state}) can be simply satisfied by initializing the system in a maximally mixed state, i.e. $\rho_{S}(0)\propto I_S$. We can prepare such an initial state by applying white noise to the qubits, since a maximally mixed state is the fixed point of random unitary transformations \cite{Burgarth:13,Novotny:09}.\\

II) Quantum annealing mode: In this mode, the Hamiltonian is time-dependent $H_S(t)$. A common example is transverse-Ising model
\begin{equation}
H_S(t)=a(t)\sum X_i+b(t)\sum J_{ij} Z_iZ_j
\end{equation}
where $Z_i$ and $X_i$ are Pauli operators and the coefficient function $a$ $(b)$ is decreasing (increasing) in time. In a hybrid, quantum-thermal approach, we want to vary the temperature during the evolution therefore we cannot reinitialize the system in a maximally mixed state as in the simulation mode. Instead, we achieve condition (\ref{initial_state}) by redefining the system-bath interaction such that $tr(\tilde{B}(t) \rho_{B}(0))=0$:

\begin{eqnarray}
H_{SB}&=&H_S+\sum_\alpha S_\alpha\otimes B_\alpha(t)+H_B\notag\\
&=&\tilde{H}_S+\sum_\alpha S_\alpha\otimes (B_\alpha(t)-\langle B_\alpha(t)\rangle)+H_B
\end{eqnarray}
This transformation modifies the system Hamiltonian $\tilde{H}_S=H_S+\sum_\alpha S_\alpha \langle B_\alpha(t)\rangle$ which as discussed below does not undermine the performance of the annealer. We consider mode II in the following device proposal. However, we consider only a time-independent system Hamiltonian $H_S$, as ideally we want to have thermalization much faster than the rate of Hamiltonian time variation in a quantum annealing protocol. Therefore we just care about the instantaneous Hamiltonian of the system.

\section{A circuit-QED Proposal}


Here we propose realizing a thermal bath by a set of driven lossy microwave resonators. Consider a system of superconducting qubits with Hamiltonian $H_S$ collectively coupled to $N_r$ resonators with the same mode frequency $\{\omega_r\}$

\begin{equation}
H_{SB}=H_S+\sum g_{\alpha \nu}S_\alpha (a_{\nu}+a^\dagger_{\nu})+\omega_r\sum_{\nu=1}^{N_r} a^\dagger_{\nu} a_{\nu} \label{collective}
\end{equation}
The system operator $S_\alpha$ is a local operator on qubit $\alpha$. In general, for each qubit we should have a coupling to the bath via two non-commuting operators $\{S^1_\alpha,S^2_\alpha\}$ so that at least one of them is non-commuting with the system Hamiltonian. We comment here that, in practice, each resonator can be coupled to a finite group of qubits. Therefore, for a large system, the bath would consist of local patches of resonators. Here for the ease of presentation we consider the case of a single collective coupling (\ref{collective}).

We classically drive each resonator mode $\omega_r$ with amplitude $\mathcal{E}_\nu$ and at frequency $\omega_d^{\nu}$ with detuning $\Delta_\nu=\omega_d^{\nu}-\omega_r$. Also each resonator has photon leakage at rate $\kappa_\nu$. The steady-state of such a driven lossy resonator is a coherent state $|\alpha_\nu\rangle=|\frac{\mathcal{E}_\nu}{\Delta_\nu+i\frac{\kappa_\nu}{2}}\rangle$. The resonators are driven in their steady-states before coupling to the system. The number of photons stored in this resonator, $\bar{N}_\nu=a_\nu^\dagger a_\nu$, fluctuates as \cite{Clerk:10}
\begin{align}
\langle \bar{N}_\nu(t)\bar{N}_\nu(t-s)\rangle-\langle \bar{N}_\nu(t)\rangle\langle\bar{N}_\nu(t-s)\rangle=e^{(i\Delta_\nu+\kappa_\nu) s}|\alpha_\nu|^2
\end{align}
with Loretzian spectrum
\begin{align}
\Lambda_\nu(\omega)=\frac{|\alpha_\nu|^2\kappa_\nu}{(\omega+\Delta_\nu)^2+\kappa_\nu^2}\label{number}
\end{align} 

Using the transformation introduced in ref. \cite{Shabani:14}, we rewrite the Hamiltonian in the dispersive regime where the resonators are detuned from system frequencies that are stronger than their coupling. Explicitly, for a system spectral decomposition $H_S=\sum_j \Omega_j |j \rangle\langle j|$, the dispersive regime is defined as 
\begin{align}
|\langle j| \sum_\alpha g_{\alpha\nu}S_\alpha |k\rangle |\ll |\omega_r-(\Omega_k-\Omega_j)|
\end{align}
Biasing the resonator in this regime requires some knowledge about the largest energy scale in the system. Assuming single qubit Pauli operators $S_\alpha$, a general but loose upper bound is
\begin{equation}
|\langle j| \sum_\alpha g_{\alpha\nu}S_\alpha |k\rangle |<\sum_\alpha g_{\alpha\nu}\ll \max_{k,j}|\omega_r-(\Omega_k-\Omega_j)|
\end{equation}

For a graph of spins coupled with Ising Hamiltonian, i.e. $H_S=\sum J_{ij}Z_iZ_j$, the energy differences $\Omega_k-\Omega_j$ can be upper-bounded by the maximum of each node's degree of connectivity times the largest edge strength, which is computable in a time linear in the number of spins. 

In this regime, as we discuss in the appendix D, the system-resonators Hamiltonian is
\begin{equation}
H^D_{SB}\approx H_S^*+\sum_{\nu} \omega_r a^\dagger_{\nu} a_{\nu}+\sum_{\nu} \hat{S}_{\nu} (a^\dagger_{\nu} a_{\nu}-\langle a_\nu^\dagger a_\nu\rangle)
\end{equation}
The system Hamiltonian is perturbatively modified as
\begin{eqnarray}
H_S^*=H_S- \sum_{\alpha,\nu}  \frac{g_{\alpha,\nu}}{2} \Big[((1+|\alpha_\nu|^2) A^\dagger_\nu-A_\nu) S_\alpha+ h.c.\Big]\label{diag-Ham}
\end{eqnarray}
with  $A_\nu=\sum_{\alpha} g_{\alpha\nu}R_\alpha$ where $R_\alpha= \sum_{jk}\frac{\langle j| S_\alpha |k\rangle }{\omega_r+\Omega_j-\Omega_k} |j\rangle\langle k|$, and system operators 
\begin{eqnarray}
\hat{S}_\nu=\frac{1}{2}\sum_{\alpha}  g_{\alpha,\nu} [S_\alpha,A_\nu^\dagger-A_\nu]
\end{eqnarray}
which yields system-resonator coupling
\begin{eqnarray}
H_I=\sum_{\alpha\beta} [S_\alpha,R_\beta^\dagger-R_\beta]\underbrace{\sum_{\nu}  \frac{g_{\alpha\nu}g_{\beta\nu}}{2}\Big(a^\dagger_{\nu} a_{\nu}-\langle a_\nu^\dagger a_\nu\rangle\Big)}_{B_{\alpha\beta}}\label{anotherone}
\end{eqnarray}

The correlation function generated by $B_{\alpha\beta}$ bath operators is

\begin{eqnarray}
&\gamma_{\alpha\beta,\alpha'\beta'}(\omega)=\frac{1}{4}\sum_{\nu=1}^{N_r}g_{\alpha\nu}g_{\beta\nu}g_{\alpha'\nu}g_{\beta'\nu}\Lambda_{\nu}(\omega)
\end{eqnarray}

In order to complete the design, the resonators should be driven such that satisfy the thermalization condition (\ref{KMS_approx}). In order to find the best parameters setting, the precision bound (\ref{precision}) suggests solving the following optimization problem

\begin{eqnarray}
\min_{N_r,\mathcal{E}_\nu,\kappa_\nu,\omega_d^\nu}\max_{\omega}&\Big| e^{\omega/T}\gamma_{\alpha'\beta',\alpha\beta}(-\omega)-\gamma_{\alpha\beta,\alpha'\beta'}(\omega)\Big|\label{precision0}
\end{eqnarray}

However, for the example presented in the following, we found the following optimization yields better numerical results
\begin{eqnarray}
\min_{N_r,\mathcal{E}_\nu,\kappa_\nu,\omega_d^\nu}\int_{\omega_{min}}^{\omega_{max}}d\omega|\frac{\gamma_{\alpha'\beta',\alpha\beta}(-\omega)}{\gamma_{\alpha\beta,\alpha'\beta'}(\omega)}- e^{-\omega/T}|\label{optimize}
\end{eqnarray}
for a given temperature $T$. Solving (\ref{optimize}) determines the design parameters: number of required resonators $N_r$ and for each resonator its drive amplitude $\mathcal{E}_\nu$, frequency $\omega_d^\nu$, and leakage rate $\kappa_\nu$. The number of resonators would be eventually limited by the fabrication constraints, while other parameters $\{\mathcal{E}_\nu,\omega_d^\nu,\kappa_\nu\}$ can be tuned on chip. Among these the only one that requires extra hardware for on-chip tuning is the leakage rate $\kappa_\nu$, for which Ref. \cite{Pierre:14} presents one particular design. In appendix C, we show that the optimization problem (\ref{optimize}) can be solved to a desired precision following the kernel properties of the Lorentzian function \cite{Barnie:10,Orlandini:10,Dirdal:2013}. 

\subsection{Qubit Chain Example}

\begin{figure}[h]
\includegraphics[width=8cm,height=2.5cm]{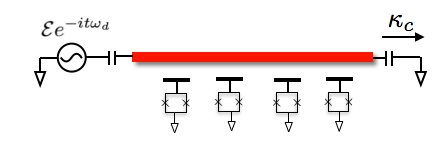}
\caption{A schematic for transmon qubits coupled to a resonator acting as a thermal bath.}\label{Xmons}
\end{figure}

\begin{figure}[h]
\includegraphics[width=7.8cm,height=5.9cm]{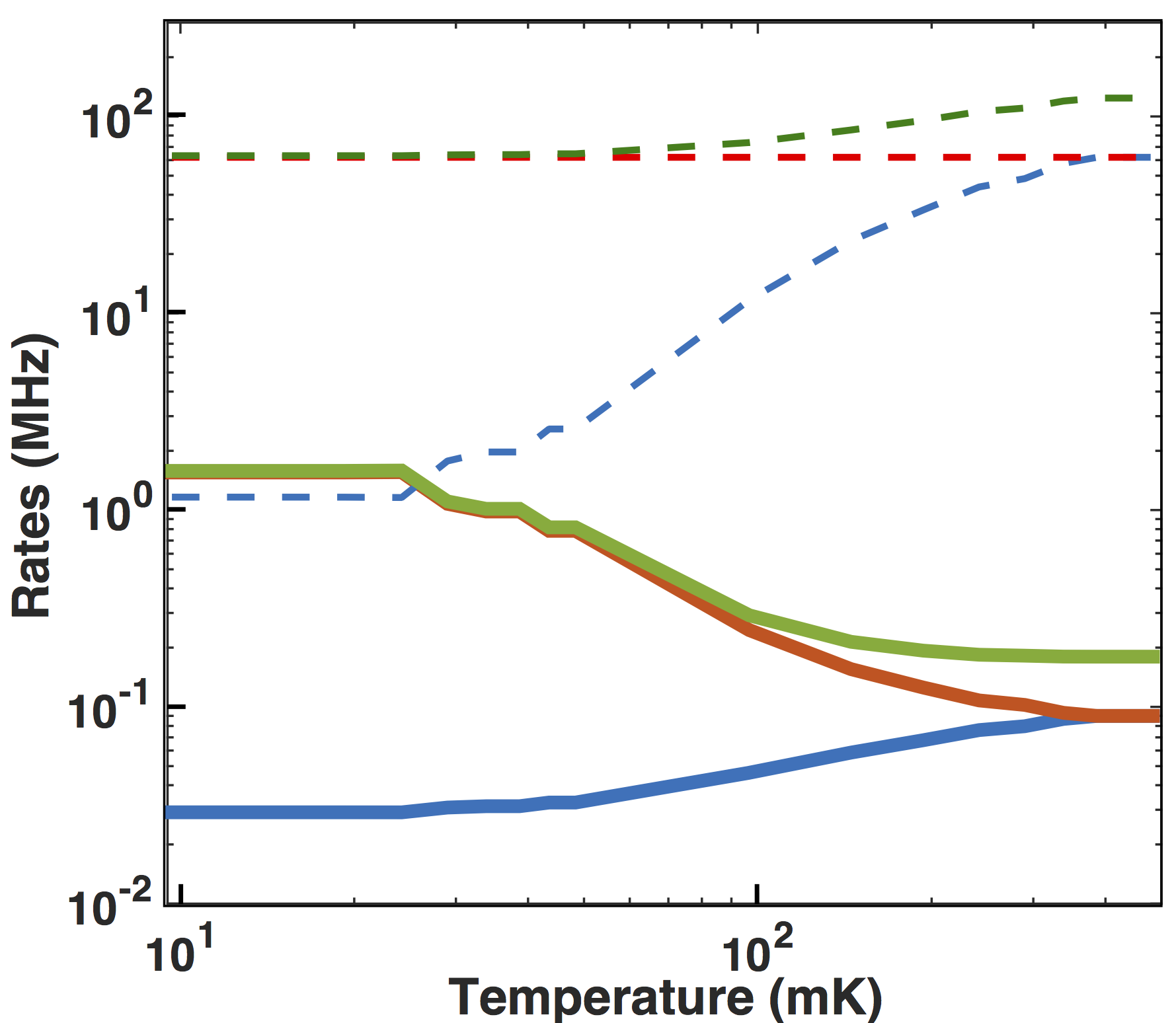}
\caption{Average rate of cooling, heating and sweeping for frequency range $[4.6,5.4]GHz$. Solid blue (red) is heating (cooling) rate for single photon drive. Dashed blue (red) is an upper bound for heating (cooling) rate. The green line corresponds to the annealing sweep rate also at the single photon level.} \label{simul}
\end{figure}

Here we discuss a basic experimental realization of the thermal bath proposal for a chain of superconducting qubits coupled to a single mode resonator. Fig.(\ref{Xmons}) is a sketch for the experiment with transmon qubits described by the Hamiltonian
\begin{eqnarray}
H_{SB}&=&H_S+\sum g_{\alpha}X_\alpha (a+a^\dagger)+ \omega_r a^\dagger a \\
H_S&=&\sum_\alpha \omega_\alpha Z_\alpha+J_{\alpha}Z_\alpha Z_{\alpha+1},
\end{eqnarray}
where $Z_\alpha$ and $X_\alpha$ are Pauli operators and real numbers $J_\alpha$ are nearest-neighbor coupling strengths. For this system, the thermal bath should operate within energy interval
\begin{eqnarray}
&&[\omega_{min},\omega_{max}]=\notag\\
&&2[\min_\alpha(\omega_\alpha- J_\alpha-J_{\alpha-1}),\max_\alpha(\omega_\alpha+ J_\alpha+J_{\alpha-1})]
\end{eqnarray}
In the numerics we consider parameters $\omega_\alpha=2.5$ GHz, $g_{\alpha}=300$ MHz, and an ensemble of random Hamiltonians with coupling $J_\alpha\in [-100,100]$ MHz.
We examine the limits of the proposal by choosing values for the resonator parameters in the range supported by the theory. Namely, the Born approximation is valid when the qubit-resonator coupling $g_\alpha^2/(\omega_r+\Delta)$, from Eq.(\ref{anotherone}), is weaker than $\omega_{min}$ and the Markovian assumption holds when the transition rates $\gamma(\omega)$ is weaker than the resonator reset rate which is the rate of photon leakage $\kappa_r$. Additionally, we need to have $\kappa_r\ll\omega_r$ for Markovian photon leakage, and $g_\alpha \ll \omega_{min}-\omega_r$ for the dispersive approximation. From master equation (\ref{Lindblad}), we find the rate of cooling and heating between levels with energy difference $\Omega$ are $\gamma(\Omega)$ and $\gamma(-\Omega)$ where
\begin{eqnarray}
\gamma(\omega)=\frac{\bar{N}\kappa_r}{(\omega+\Delta)^2+(\frac{\kappa_r}{2})^2}(\frac{2\omega g^2}{\omega_r^2-\omega^2})^2 \label{rate_eq}
\end{eqnarray}
Here $\bar{N}$ is the average photon number stored in the resonator, which is proportional to the incident power. The only parameter that we optimize in condition (\ref{optimize}) is the drive detuning $\Delta$ such that the Gibbs state is prepared with above $95\%$ fidelity. We use MATLAB's fmincon function to perform this optimization. All drives are red detuned, i.e. $\Delta<0$. Fig.(\ref{simul}) shows the average rate of cooling and heating for range $\Omega\in[\omega_{min},\omega_{max}]$. The solid blue and red lines are rates of heating and cooling per single photon $\bar{N}=1$ for different target temperatures. As we drive the resonator more strongly, the rates increase. Plots of dotted blue and red lines show the maximum possible rates before the perturbative assumptions of Eq.(\ref{Lindblad}) fail. Specifically, we consider the limit setting of $\omega_r=\omega_{min}-5g_\alpha$ and $\kappa_r=0.2 \omega_r$ which gives the values $\omega_r=3.1 GHz$ and $\kappa_r=620 MHz$. Then we choose the maximum allowed cooling rate (the straight line upper bound) to be $\gamma(\Omega)=62 MHz$, a factor of $10$ smaller than $\kappa_r$ in order to ensure the Markovianity of dynamics. This is just a crude estimate. In practice, the resonator enters its regime of nonlinearity for strong drives, restricting the maximum achievable heating and cooling rates. We also plot in green the summation of heating and cooling rate that corresponds to the all-spin sweep rate in simulated thermal annealing \cite{Boixo:2014}. Since the Hamiltonian is Ising and bath operations are spin flips, within current technology the relevant imperfection is the presence of dissipation with a $T_1$ time scale of about $6.2$ kHz \cite{dial:2015}. 

As explained in appendix D, one perturbative effect due to dispersive corrections is the presence of an effective decoherence similar to the Purcell effect. This effect is negligible in the regime of parameters for the above example, therefore we ignore it as we expect small corrections to the plots in fig.(\ref{simul}). However, this factor may need to be accounted for other physical scenarios as it perturbs the final equilibrium state of the system.

\section{Limitations of the circuit-QED proposal}

Here we discuss some potential limitations of the above proposal from hardware and application perspective. In a device implementation, we would be able to couple a qubit to a finite number of resonators and therefore strictly speaking, this proposal is not scalable. We examined that driving resonators with a multi-frequency input rather than a single-frequency field does not help reducing the number of required number of resonators $N_r$. The reason is that a multi-frequency drive yields a time-dependent spectrum making it nontrivial to satisfy the thermalization condition (\ref{thermal_rondition}). We expect a scalable design would be possible with meta-material with a continuum of modes \cite{Egger:13,Haeberlein:15}, instead of a discrete set of resonators. However, the current theory for meta-materials (both design and dynamical response) is very limited and requires fundamental developments before we can apply it to design a thermal bath. 

The rate of thermalization, given finite coherence times, is an important factor for any useful artificial thermal bath proposal. In section III.A, we numerically evaluated the thermalization rate for a one-dimensional chain problem. This rate, in general, can be increased by raising the strength of system-bath coupling, however in the above formulation, increasing $g_{\alpha\nu}$, also perturbs the system Hamiltonian due to the dispersive corrections. Specifically, both modifications to the system Hamiltonian $||H_S^*-H_S||_1$ and system-bath couplings are proportional to $\max|\frac{g_{\alpha\nu}g_{\beta\nu}}{\omega_r+\Omega_j-\Omega_k}|$. In the context of annealing, this Hamiltonian perturbation is not a problem since we turn off the system-bath coupling towards the end of annealing. However, this Hamiltonian perturbation becomes a limitation in simulation mode. In this mode, system equilibrates at the state $\frac{e^{-H^*_S/T}}{{\rm{Tr}}(e^{-H^*_S/T})}$ instead of $\frac{e^{-H_S/T}}{{\rm{Tr}}(e^{-H_S/T})}$. This error in state preparation can be related to Hamiltonian modification with
\begin{eqnarray}
||\frac{e^{-\beta H_1}}{Tr(e^{-\beta H_1})}-\frac{e^{-\beta H_2}}{Tr(e^{-\beta H_2})}||_1\leq 2(e^{\beta ||H_1-H_2||_1}-1)
\end{eqnarray}
which means exponential sensitivity, and therefore a limitation of our proposal to achieve high fidelities for low temperatures simulations. There is a trade-off between the accuracy of simulations versus rate of equilibration that is inversely proportional to the total simulation time. A technical note is that, we used Von Neumann's trace inequality to derive this bound \cite{Mirsky:75}. 

\section{A Bath of Superconducting Qubits}

In the earliest proposal for realizing a quantum thermal state, Ref. \cite{Terhal:02}, Terhal and Divincenzo suggested a simple way to engineer the bath: preparing a number of ancilla qubits in their thermal state and repetitively interacting them with the main system for a short period. In this scheme, after each time of interaction the ancilla need to be reinitialized in their thermal state. The corresponding dynamics would be described with a thermalizing Lindblad equation (\ref{Lindblad}). This proposal has certain limitation that make it challenging for practical purposes. A qubit's initialization in a thermal state can be a relatively slow process as it would involve a non-unitary dissipative dynamics, for instance tens of nanoseconds for superconducting qubits. Therefore a few iterations would take longer than the coherence time of the qubits. It would be an interesting topic of research to resolve this limitation by engineering fast qubit reset dynamics or tunable strongly resistively elements.

\section{Conclusion}
Engineering a quantum simulator to simulate the thermodynamical properties of a many-body quantum system has been a fundamental problem in quantum computation. In this paper, we propose an analog implementation of a quantum simulator that mimics a thermalization process. In contrast to the quantum Metropolis algorithm that requires a universal fault-tolerant quantum computer, our analog proposal allows an approximate simulation on a midsize quantum system. We also discussed that an engineered thermal bath has applications in quantum annealing. A quantum annealing processor with an artificial temperature knob allows running annealing process at non-zero temperatures in a controllable fashion. In this paper, we examined our proposal for a simple one-dimensional spin system. Finally, we should emphasize that this is paper intended to present the physics principles and not an ultimate hardware design. However, any modified version of this proposal would follow the principle of a universal thermal bath defined in this paper.

\section{Acknowledgments}
We thank Kristan Temme for helpful discussions on quantum Metropolis algorithms.

\section{Appendices}


\subsection{Proof of Thermalization Condition}

In order to prove that Eq. (7) suffices for the Gibbs state to be a fixed point, we need to show $\mathcal{D}(\rho_{th})=0$ holds assuming (7). Using the identities: $\rho_{th}S_\alpha(\omega)=e^{\omega/T}S_\alpha(\omega)\rho_{th}$ and $S_\alpha^\dagger(\omega)=S_\alpha(-\omega)$, we find
\begin{widetext}
\begin{eqnarray}
\mathcal{D}(\rho_{th})&=&\sum_{\omega>0,\alpha,\alpha'}\gamma_{\alpha\alpha'}(\omega,t)\Big(e^{-\omega/T}S_{\alpha'}(\omega) S_\alpha(-\omega)-S_\alpha(-\omega)S_{\alpha'}(\omega)\Big)\rho_{th}\notag\\
&&+\gamma^*_{\alpha\alpha'}(\omega,t)\Big(e^{-\omega/T}S_{\alpha}(\omega)S_{\alpha'}(-\omega)-S_{\alpha'}(-\omega)S_{\alpha}(\omega)\Big)\rho_{th}\notag\\
&&+\gamma_{\alpha\alpha'}(-\omega,t)\Big(e^{\omega/T}S_{\alpha'}(-\omega) S_\alpha(\omega)-S_\alpha(\omega)S_{\alpha'}(-\omega)\Big)\rho_{th}\notag\\
&&+\gamma^*_{\alpha\alpha'}(-\omega,t)\Big(e^{\omega/T}S_{\alpha}(-\omega)S_{\alpha'}(\omega)-S_{\alpha'}(\omega)S_{\alpha}(-\omega)\Big)\rho_{th}\notag
\end{eqnarray}
Next we use thermal condition $\gamma_{\alpha\alpha'}(-\omega,t)=e^{-\omega/T}\gamma_{\alpha'\alpha}(\omega,t)$:
\begin{eqnarray}
\mathcal{D}(\rho_{th})&=&\sum_{\omega>0,\alpha,\alpha'}\gamma_{\alpha\alpha'}(\omega,t)\Big(e^{-\omega/T}S_{\alpha'}(\omega) S_\alpha(-\omega)-S_\alpha(-\omega)S_{\alpha'}(\omega)\Big)\rho_{th}\notag\\
&&+\gamma^*_{\alpha\alpha'}(\omega,t)\Big(e^{-\omega/T}S_{\alpha}(\omega)S_{\alpha'}(-\omega)-S_{\alpha'}(-\omega)S_{\alpha}(\omega)\Big)\rho_{th}\notag\\
&&+\gamma_{\alpha'\alpha}(\omega,t)\Big(S_{\alpha'}(-\omega) S_\alpha(\omega)-e^{-\omega/T}S_\alpha(\omega)S_{\alpha'}(-\omega)\Big)\rho_{th}\notag\\
&&+\gamma^*_{\alpha'\alpha}(\omega,t)\Big(S_{\alpha}(-\omega)S_{\alpha'}(\omega)-e^{-\omega/T}S_{\alpha'}(\omega)S_{\alpha}(-\omega)\Big)\rho_{th}\notag
\end{eqnarray}

which shows that this last summation is also zero.
\end{widetext}

Now we consider the uniqueness of the fixed point which is guaranteed by ergodicity of the dynamics. The Davies-Frigiero-Spohn criterion \cite{Frigerio:78} states that given Lindblad dynamics with diagonal form $\dot{\rho}=-i[H,\rho]+\sum_j 2L_j\rho L_j^\dagger-L_j^\dagger L_j\rho-\rho L_j^\dagger L_j$, the necessary and sufficient condition for the uniqueness of a steady-state is $\{L_{j},L^\dagger_{j},H_S\}^{'}=c\mathbb{1}$, where $\{.\}^{'}$ indicates the commutant and $c$ is a constant. The Lindblad equation (5) is not in diagonal form, and in the lab frame $\dot{\rho}=-i[H_S,\rho]+\sum_{\omega,\alpha,\alpha'}\gamma_{\alpha\alpha'}(\omega,t)\Big(2S_{\alpha'}(\omega)\rho S^\dagger_\alpha(\omega)-\{S^\dagger_\alpha(\omega)S_{\alpha'}(\omega),\rho\}\Big)$. Here we consider only time-homogenous dynamics. The coefficients $\gamma_{\alpha\alpha'}$ change continuously as we change the effective temperature, meanwhile the Davies-Frigiero-Spohn criterion for the diagonalized form of our Lindblad equation should hold for any values of $\gamma_{\alpha\alpha'}$. In this situation, following some simple linear algebra, the necessary and sufficient condition for uniqueness of Gibbs state as a fixed point is
\begin{eqnarray}
\{S_{\alpha}(\omega),S^\dagger_{\alpha}(\omega),H_S\}^{'}=c\mathbb{1}
\end{eqnarray}
We can justify this condition as follows: for an element $T$ of the commutant set we have $[T,H_S]=0$, ($H_S=\sum_j\Omega_j |j\rangle\langle j|$), therefore $T=\sum_j t_j |j\rangle\langle j|$ for some numbers $t_j$. $T$ should also commute with $S_{\alpha}(\omega)$, so that
\begin{eqnarray}
[T,S_\alpha(\omega)]=[\sum_j t_j |j\rangle\langle j|,\sum_{\epsilon'-\epsilon=\omega}\Pi(\epsilon)S_\alpha\Pi(\epsilon')]\notag\\
=\sum_{\epsilon'-\epsilon=\omega}(t_{\epsilon}-t_{\epsilon'})\Pi(\epsilon)S_\alpha\Pi(\epsilon')=0
\end{eqnarray}
This dictates a block-diagonal form of each proportional to identity, $T=\oplus_q c_q I_q$. The unique fixed point condition is satisfied if $T$ reduces to a single block or $c_q=c$ $\forall q$. Such a condition holds for any two energy levels $\{|\epsilon\rangle,|\epsilon'\rangle\}$ iff either there exist operators $\{S_{\alpha_j}\}$ and intermediate levels $\{|\epsilon_j\rangle\}$ such that either $\langle\epsilon|S_\alpha|\epsilon'\rangle\neq 0$ or $\langle\epsilon|S_{\alpha_j}|\epsilon_j\rangle ... \langle\epsilon_{j'}|S_{\alpha_{j'}}|\epsilon'\rangle\neq 0$. This has a simple physical interpretation, which is that population can be transferred between all levels via coupling to the bath.

\subsection{Precision of the Approximate Thermalization Condition}

In practice any engineered thermal bath can only approximately satisfy the thermalization condition: 
 
 \begin{eqnarray}
\frac{\gamma_{\alpha\alpha'}(-\omega)}{\gamma_{\alpha'\alpha}(\omega)}\approx e^{-\omega/T} \label{KMS_approx_app}
\end{eqnarray}
therefore we need to address the required precision in this approximation. 
We use the result in Ref. \cite{Szehr:13} on the stability of quantum Markov processes to derive a bound on the error of Gibbs state preparation $||\rho_{eq}-\rho_{th}||_1$, here $\rho_{eq}$ is the equilibrium  engineered state and $\rho_{th}=\frac{e^{-H_S/T}}{{\rm{Tr}}(e^{-H_S/T})}$ is the target thermal state. Here we use trace norm for the operator where $||X||_1=tr[\sqrt{X^\dagger X}]$ is the trace norm. The norm that we use for operator maps is 
\begin{equation}
||\mathcal{Q}||_{1\rightarrow 1}:=\sup_{X\in {\mathcal{M}}_r(\mathbb{C})} \frac{||\mathcal{Q}(X)||_1}{||X||_1}
\end{equation}

We first express the following theorem as the result of theorem 6 and corollary 7 from Ref. \cite{Szehr:13}.

{\theorem{Consider two Lindblad maps $\mathcal{L}_1,\mathcal{L}_2$ acting on a $d$-dimensional Hilbert space, ${\mathcal{M}}_d(\mathbb{C})$. Furthermore assume $\mathcal{L}_1$ has a unique fixed point corresponding to the zero eigenvalue, and the first non-zero eigenvalue $\lambda$ determines the rate of equilibration as $||\mathcal{T}^t-\mathcal{T}^\infty||_{1\rightarrow 1}\leq K e^{-\lambda t}$. Here $K<d$ is a constant. If $\rho_1$ is the unique fixed state of $\mathcal{L}_1$ and $\rho_2$ is the fixed point of the map $\mathcal{L}_2$, then we find the following bounded distance between the fixed points of the two map $\mathcal{L}_1$ and $\mathcal{L}_2$}}

\begin{equation}
||\rho_1-\rho_2||_1\leq\frac{\log(d)+1}{\lambda} ||\mathcal{L}_1-\mathcal{L}_2||_{1\rightarrow 1}\label{bound}
\end{equation}

We use this theorem to derive the robustness of the approximate thermalization condition. Suppose we engineer a thermal bath approximately satisfying the thermalization condition (\ref{KMS_approx_app})

\begin{eqnarray}
\frac{d\tilde{\rho}_S}{dt}&=&\mathcal{D}(\tilde{\rho}_S)=\sum_{\omega,\alpha,\alpha'}\gamma_{\alpha\alpha'}(\omega)\times\notag\\
&&\Big(S_{\alpha'}(\omega)\tilde{\rho}_S S^\dagger_\alpha(\omega)-\frac{1}{2}\{S^\dagger_\alpha(\omega)S_{\alpha'}(\omega),\tilde{\rho}_S\}\Big) \label{Lindblad2}
\end{eqnarray}
with the fixed point $\rho_{eq}$ . Now we construct a master equation that satisfies the exact thermalization condition and therefore has a fixed point $\rho_{th}=\frac{e^{-H_S/T}}{{\rm{Tr}}(e^{-H_S/T})}$:

\begin{eqnarray}
\frac{d\tilde{\rho}_S}{dt}&=&\mathcal{D}^*(\tilde{\rho}_S)=\sum_{\omega,\alpha,\alpha'}\gamma^*_{\alpha\alpha'}(\omega)\times\notag\\
&&\Big(S_{\alpha'}(\omega)\tilde{\rho}_S S^\dagger_\alpha(\omega)-\frac{1}{2}\{S^\dagger_\alpha(\omega)S_{\alpha'}(\omega),\tilde{\rho}_S\}\Big) \label{Lindblad3}
\end{eqnarray}
where $\gamma^*_{\alpha\alpha'}(\omega)=\gamma_{\alpha\alpha'}(\omega)$ for $\omega\geq 0$, and $\gamma^*_{\alpha'\alpha}(\omega)=\gamma_{\alpha'\alpha}(-\omega)e^{\omega/T}$ for $\omega<0$. We find the distance between Markovian semigroup generators $\mathcal{D}$ and $\mathcal{D}^*$
\begin{widetext}
\begin{eqnarray}
||\mathcal{D}^*-\mathcal{D}||_{1\rightarrow 1}= ||\sum_{\omega<0,\alpha,\alpha'}\Big( \gamma^*_{\alpha\alpha'}(\omega)-\gamma_{\alpha\alpha'}(\omega)\Big)\Big(S_{\alpha'}(\omega){\bf{.}} S^\dagger_\alpha(\omega)-\frac{1}{2}\{S^\dagger_\alpha(\omega)S_{\alpha'}(\omega),{\bf{.}}\}\Big)||_{1\rightarrow 1}\notag\\
\leq \sum_{\omega<0,\alpha,\alpha'}\Big| \gamma^*_{\alpha\alpha'}(\omega)-\gamma_{\alpha\alpha'}(\omega)\Big|\times||S_{\alpha'}(\omega){\bf{.}} S^\dagger_\alpha(\omega)-\frac{1}{2}\{S^\dagger_\alpha(\omega)S_{\alpha'}(\omega),{\bf{.}}\}||_{1\rightarrow 1}\label{D-bound}
\end{eqnarray}
\end{widetext}
Using the operator norm inequalities, we find
\begin{eqnarray}
||S_{\alpha'}(\omega){\bf{.}} S^\dagger_\alpha(\omega)-\frac{1}{2}\{S^\dagger_\alpha(\omega)S_{\alpha'}(\omega),{\bf{.}}\}||_{1\rightarrow 1}\notag\\
\leq 2||S_{\alpha}(\omega)||_{1\rightarrow 1}||S_{\alpha'}(\omega)||_{1\rightarrow 1}
\end{eqnarray}
The form of $S_\alpha(\omega)$ was given in the paper $S_\alpha(\omega)=\sum_{\epsilon'-\epsilon=\omega}\Pi(\epsilon)S_\alpha\Pi(\epsilon')$. Consider system energy gaps have degeneracy $Gen(\omega)$ that is basically the number of term in the summation $\sum_{\epsilon'-\epsilon=\omega}$. Note that $Gen(\omega)$ is automatically zero when $\Pi(\epsilon)S_\alpha\Pi(\epsilon')=0$ and that is the case when the system operator makes no transition between states $|\epsilon\rangle$ and $|\epsilon'\rangle$.Then we have
\begin{eqnarray}
||S_{\alpha}(\omega)||_{1\rightarrow 1}\leq Gen(\omega) \max_{\epsilon'-\epsilon=\omega}|\langle\epsilon|S_\alpha|\epsilon'\rangle|^2
\end{eqnarray}
where $\Pi(\epsilon)=|\epsilon\rangle\langle\epsilon|$ and $\Pi(\epsilon')=|\epsilon'\rangle\langle\epsilon'|$. If we further consider the general model of single qubit coupling to the bath, $S_\alpha\in\{X,Y,Z\}$, we find $||S_{\alpha}(\omega)||_{1\rightarrow 1}\leq Gen(\omega)$. Plugging this into Eq. (\ref{D-bound})
\begin{widetext}
\begin{eqnarray}
||\mathcal{D}^*-\mathcal{D}||_{1\rightarrow 1}\leq 2\sum_{\omega<0,\alpha,\alpha'}Gen^2(\omega)\Big| \gamma^*_{\alpha\alpha'}(\omega)-\gamma_{\alpha\alpha'}(\omega)\Big|\leq 6\Big(\sum_{\omega<0}Gen(\omega)\Big )^2\max_{\omega<0,\alpha,\alpha'}\Big| \gamma^*_{\alpha\alpha'}(\omega)-\gamma_{\alpha\alpha'}(\omega)\Big|
\end{eqnarray}
\end{widetext}
The summation $Gen(\omega)$ is simply the total number of energy level transitions corresponding to a nonzero $\langle\epsilon|S_\alpha|\epsilon'\rangle$. For an arbitrary $n$-qubit system Hamiltonian $\sum_{\omega<0}Gen(\omega)=\frac{d(d-1)}{2}=2^{2n-1}-2^{n-1}$. However, for an Ising system Hamiltonian and $S_\alpha=X$, $\langle\epsilon|S_\alpha|\epsilon'\rangle$ is non-zero only for $|\epsilon\rangle$ and $|\epsilon'\rangle$ with only a single bit flip distance. In this case $\sum_{\omega<0}Gen(\omega)=\frac{\log^2(d)+\log(d)}{2}=\frac{n(n+1)}{2}$. Now we can relate the precision in thermal bath preparation with the accuracy of the thermalization condition. 
\begin{widetext}
\begin{eqnarray}
||\rho_{eq}-\rho_{th}||_1&\leq& 6\frac{\log(d)+1}{\lambda} \Big(\sum_{\omega<0}Gen(\omega)\Big )^2\max_{\omega<0,\alpha,\alpha'}\Big| \gamma^*_{\alpha\alpha'}(\omega)-\gamma_{\alpha\alpha'}(\omega)\Big|\\
&=& 6\frac{\log(d)+1}{\lambda} \Big(\sum_{\omega<0}Gen(\omega)\Big )^2\times\max_{\alpha,\alpha',\omega\in [-\omega_{max},-\omega_{min}]}\Big| e^{\omega/T}\gamma_{\alpha'\alpha}(-\omega)-\gamma_{\alpha\alpha'}(\omega)\Big|\notag
\end{eqnarray}
\end{widetext}

\subsection{Approximating with a Sum of Lorentzians}

In order to argue that a set of resonators with Lorentzian fluctuation functions can approximate the thermalization condition, we consider the following scenario. Split the resonators in two groups $a$ and $b$ and for simplicity consider qubit-resonator couplings have the same strength, i.e. $g_{\alpha\nu}=g_{\beta\nu}=g$, then the collective spectrum is
\begin{eqnarray}
\gamma^a(\omega)=\frac{g^4}{4}\sum_{\nu\in a}\Lambda_{\nu}(\omega), \gamma^b(\omega)=\frac{g^4}{4}\sum_{\nu\in b}\Lambda_{\nu}(\omega) 
\end{eqnarray}
Consider the following particular way to satisfy the thermalization condition by setting
\begin{eqnarray}
\gamma^a(\omega)\approx e^{\omega/T}F(\omega),  \gamma^b(\omega)\approx 0  \label{KMS1}
\end{eqnarray}
for $\omega\in[-\omega_{max},-\omega_{min}]$ and
\begin{eqnarray}
\gamma^a(\omega)\approx 0, \gamma^b(\omega)\approx F(\omega)  \label{KMS2}
\end{eqnarray}
for $\omega\in[\omega_{min},\omega_{max}]$, and for some even function $F(\omega)$. Since any analytic function can be approximated arbitrarily accurately with a sum of Lorentzian functions \cite{Dirdal:2013}, we can tune bath mode groups $a$ and $b$ independently such that relations (\ref{KMS1}) and (\ref{KMS2}) are simultaneously satisfied, and thus the thermalization condition. Note that this is mere a mathematical argument and not necessarily the way to use resonators in a real implementation. Solving optimizations (\ref{precision0}) or (\ref{optimize}) is what should be considered in a design.

\subsection{Dispersive Regime Discussion}

The detailed derivation of the modifications due to dispersive transformation is given in Ref. \cite{Shabani:14}. Start with the Hamiltonian in Eq.(12)

\begin{equation}
H_{SB}=H_S+\sum g_{\alpha \nu}S_\alpha (a_{\nu}+a^\dagger_{\nu})+\sum_{\nu} \omega_r a^\dagger_{\nu} a_{\nu} \label{collective-app}
\end{equation}
Apply the dispersive transformation $U_{D}=\exp[\sum A_\nu a_\nu^\dagger -A_\nu^\dagger a_\nu]$, where $A_\nu=\sum_{jk} \frac{|\langle j| \sum_\alpha g_{\alpha\nu}S_\alpha |k\rangle |}{\omega_r+\Omega_j-\Omega_k} |j\rangle\langle k|$. Assuming different modes are driven in off-detuned frequencies we drop the cross-mode coupling terms $a_\nu^\dagger a_{\nu'}$ induced by the dispersive transformation and arrive at

\begin{equation}
H^D_{SB}=U_DH_{SB}U_D^\dagger\approx H_S^D+\sum_{\nu} \omega_r a^\dagger_{\nu} a_{\nu}+\sum_{\nu} S_{\nu} a^\dagger_{\nu} a_{\nu}\label{Hsc}
\end{equation}
where the modified system Hamiltonian is $H_S^D=H_S-\frac{1}{2} \sum_{\alpha,\nu}  g_{\alpha,\nu} (A_\nu^\dagger S_\alpha+ S_\alpha A_\nu)$, and the new system operators for resonator couplings are $S_\nu=\frac{1}{2}\sum_{\alpha}  g_{\alpha,\nu} [S_\alpha,A_\nu^\dagger-A_\nu]$. Cavity leakage causes additional system decoherence (Purcell-like effect). Following Eq.(4) in Ref. \cite{Shabani:14}, it is caused by a virtual coupling between the system and the outside reservoir. The corresponding decoherence rate is $\kappa_\nu\frac{g_{\alpha,\nu}^2}{(\omega_r-(\Omega_j-\Omega_k))^2}$, which we can ignore in comparison with the heating rate Eq.(25) when $\bar{N}>\frac{1}{4}(1+\frac{\Delta}{\Omega})$.
The drive Hamiltonians are also modified under the dispersive transformation: the system drive term at frequency $\omega_r^d$ is far off-resonance so that it does not excite the system, and this still holds when corrections of order $\frac{g_{\alpha,\nu}}{\omega_r-(\Omega_j-\Omega_k)}$ are added.

\end{document}